\documentclass[11pt,a4paper]{article}
\usepackage[hyperref]{emnlp2020}
\usepackage{times}
\usepackage{latexsym}
\usepackage{graphicx}

\usepackage{microtype}

\aclfinalcopy 

\def\mymathhyphen{{\hbox{-}}}

\title{ELSKE: Efficient Large-Scale Keyphrase Extraction}

\author{Johannes Knittel, Steffen Koch, Thomas Ertl \\
  Institute for Visualization and Interactive Systems \\
  University of Stuttgart \\
  \small\texttt{firstname.lastname@vis.uni-stuttgart.de}}

\date{}

\begin{document}
\maketitle
\begin{abstract}
Keyphrase extraction methods can provide insights into large collections of documents such as social media posts.
Existing methods, however, are less suited for the real-time analysis of streaming data, because they are computationally too expensive or require restrictive constraints regarding the structure of keyphrases.
We propose an efficient approach to extract keyphrases from large document collections and show that the method also performs competitively on individual documents.
\end{abstract}

\section{Introduction}

Automatically extracting descriptive words (keywords) or phrases (keyphrases) from documents is important for a wide range of tasks, including document summarization and improved information retrieval in databases~\cite{AlamiMerrouni2020}.
Several graph-~\cite{Mihalcea2004,Wan2008,Bougouin2013,Skrlj2019}, statistical-~\cite{El-Beltagy2009,Rose2010,Campos2020}, and machine learning-based methods~\cite{Meng2017,Xiong2019,Wang2019,Ye2020,Santosh2020} have been developed to find a limited set of concise words or phrases that best describe a certain document.

A typical keyphrase extraction pipeline consists of two steps~\cite{Hasan2010}.
First, the algorithm extracts a set of candidate phrases.
Then, a suitable ranking is applied to retrieve the best fits.
Part-of-Speech tagging is often used to retrieve candidates that are composed of nouns and adjectives~\cite{Hasan2010,Mihalcea2004,Wan2008}, but this excludes longer sequences.
Furthermore, most POS taggers need a considerable amount of processing time~\cite{Horsmann2015}.
YAKE~\cite{Campos2020} does not make use of POS tagging, but focuses on extracting up to tri-grams per default.
In recent years, machine learning techniques have been proposed that significantly outperformed previous state of the art, but powerful models are computationally expensive, need extensive training data, and may generalize less to foreign domains due to the supervised training.

In this work, we shift the focus slightly from documents to collections of (micro-)documents.
Apart from analyzing individual documents, keyphrase extraction methods can also provide insights into large document collections, for instance, to gain an overview of recent news reports or trending topics on social media.
Existing methods largely focus on single documents and do not take into account the particular challenges of analyzing streaming data such as continuously incoming tweets.
Short documents provide little context that can be harvested for mining descriptive keyphrases, and the sheer quantity of newly published items per second requires a great amount of computational resources or efficient methods.
This is particularly important for applications that need to process incoming documents immediately, e.g., in scenarios that aim at providing situational awareness.
Dealing with large collections, we may also want to find frequent longer phrases to better understand the underlying data with additional context information.

Unfortunately, we can hardly rely on syntactical and structural assumptions regarding phrase candidates if we need to avoid Part-of-Speech tagging for efficiency.
Extracting context-rich keyphrases from large datasets in a timely manner is therefore particularly challenging. 
The complexity of extracting every possible $n$-gram increases linearly with $n$, and an extended set of candidates will also lead to an increase of similarly worded keyphrases.

Hence, we propose a new method to efficiently extract descriptive, but potentially long phrases that appear unusually often, including complete sentences.
For the ranking, we extend the concept of TF-IDF to phrases and adapt it to the analysis of large document collections.

\section{Method}
We call the document or the collection of documents from which we want to extract keyphrases \emph{source}.
This means that if we want to analyze collections, we concat the individual documents into one big document.
Let $V$ be our vocabulary of terms in the source, then each keyphrase $p^i$ is a sequence $(v_1^i, …, v_m^i)$ with $v_j^i \in V$.
While TF-IDF to rank candidate keyphrases is often used as a baseline to evaluate more advanced ranking approaches, it performs surprisingly well in combination with Part-of-Speech tagging~\cite{Hasan2010,Meng2017} and, importantly, has little requirements and external dependencies.
The main idea of TF-IDF for ranking keywords is to weight terms $v^i$ with their frequency in the source $f_s(v^i)$ in relation to the document frequency of the term in a reference collection $f_d(v^i)$ comprising $N$ documents:

\begin{equation}
\mathrm{TF \mymathhyphen IDF}(v^i) = f_s(v^i)  \ln{ \frac{N}{f_d(v^i)}}
\end{equation}

A list of stop words that should be ignored is often used to greatly improve the results.
One way to extend TF-IDF to phrases is to sum up the individual scores of each term~\cite{Hasan2010}, but this favors long phrases.
In this work, we therefore set the \emph{phrase frequency} in the source in relation with the document frequency of the phrase in a reference collection.
Unfortunately, with an increasing number of words in the source the relation between those two components diverge and the influence of the inverse document frequency diminishes.
For typical English documents, e.g., news reports, the maximum term frequency ranges around 500.
If we analyze the concatenation of thousands or even hundreds of thousands of documents, however, the most frequent term can easily appear more often than 10 000 times.
Hence, we introduce a sublinear scaling factor to adapt the phrase frequency depending on the size of the source:

\begin{equation}
\mathrm{PF \mymathhyphen IDF}(p^i) = s(p^i)= f_s(p^j)^{\frac{1}{\mu}}  \ln{ \frac{N}{f_d(p^j)}}
\end{equation}

If the maximum term frequency $f_s^{\mathrm{max}}$ exceeds $500$, we set $\mu = \log_{500}{f_s^{\mathrm{max}}}$, otherwise $\mu=1$.
This means we non-linearly scale the maximum term frequency in the source down to the upper limit of $500$ while keeping the scaled frequency of terms that only appear once at $1$.
In other words, we adjust the term or phrase frequency such that the typical relation between term frequency and inverse document frequency remains similar irrespectively of the size of the collection.

\subsection{Efficiently Extracting Candidate Phrases}

Unfortunately, extracting every $\{1,2, …, m\}$-gram in the source and computing the corresponding document frequency is not feasible, especially if the source comprises millions of sentences.
We exploit the fact that in most use cases we only want to extract the top $k$ keyphrases, e.g., the top 1000.
With this assumption we can speed up the process of extracting candidates as described in this section.

\paragraph{(1) Extracting Uni- And Bigrams:}

We first extract uni- and bigrams (excluding stop words), calculate their respective PF-IDF score $s(p^i)$, sort the results in descending order, and store the source and document frequencies in a map-like structure for fast retrieval.
The score at position $k$ ($s_k$) is a lower limit, i.e., we can divide $s_k$ by the maximum possible inverse document frequency $\log{N}$ and raise it to the power of $\mu$ to retrieve the minimum frequency threshold $f^{\mathrm{th}}$.
It follows that every pattern in our final top $k$ has to have a frequency of at least $f^{\mathrm{th}}$.
Hence, we only need to extract phrases that appear at least as often as our threshold in the source.
The higher $f^{\mathrm{th}}$, the higher the speedup.

\paragraph{(2) Extracting Longer Phrases:}

We now need to calculate the frequencies of phrases that contain more than two words.
We ignore phrases that only contain stop words or appear only once (in case the minimum frequency is 1).
In the naive approach we would need to look at and count every $\{1,2, …, m\}$-gram at every position in the source.
With the lower limit of $f^{\mathrm{th}}$, though, we can stop the inner loop earlier if the frequency of the current bigram is below the threshold.
The frequency of any bigram in a sequence is an upper limit of the frequency of that sequence and any longer sequence.
Retrieving the bigram frequency is in $O(1)$ because we have already counted these in the first step.
At the end of this step we discard every phrase that does not meet our frequency threshold of $f^{\mathrm{th}}$.

\paragraph{(3) Discarding Redundant Sub-Phrases:}

For each phrase $(v_1^i, …, v_m^i)$ we also have $m-1$ sub-phrases ${ (v_1^i), (v_1^i, v_2^i),...}$ in our candidate set from the previous steps.
We want to discard such sub-phrases that have the same frequency in the source, because they only appear as part of the longer sequence.

\paragraph{(4) Calculating The Document Frequency:}

We need to retrieve the document frequency of each candidate phrase in our reference collection.
To speed up the process we can first build a bigram-based index of the collection.
Then, we can calculate the PF-IDF score of every phrase and discard patterns with a lower score than our threshold $s_k$.

\subsection{Condensing The Set Of Candidates}

After applying the first part of our pipeline, we could already extract the top k $\{1,2, ...,m\}$-grams in the source according to the PF-IDF weighting scheme.
However, among these candidates we often have several variations of similar phrases with slightly different frequencies. Hence, we want to further condense the field of candidates to retrieve the most salient and descriptive keyphrases.

\paragraph{Stop Word-Heavy Candidates:}
We first remove candidates if they only contain one term $v_i$ that is not a stop word and $s(v_i) < s_k$, i.e, only the additional stop word put the term above the threshold.

\paragraph{Redundant Longer Candidates:}
Second, we remove longer phrases that provide little additional context.
For instance, we want to discard \textit{at a birthday party} if \textit{birthday party} is one of our candidate phrases.
We say a phrase $p^j$ is a longer phrase of $p^i$ if the sequence $p^j$ contains the sequence $p^i$.
For each phrase $p^i$ we determine whether there is a longer phrase $p^j$ with at most two additional words in front of $p^i$ and/or after $p^i$, the \emph{overhang}.
We only keep the longer phrase if the individual PF-IDF score of any overhang is high enough (and does not contain only stop words), i.e., $s(v_j) \geq \lambda s_k$ for an overhanging word $v_j$ or $s((v_j, v_{j+1})) \geq \lambda s_k$ for an overhanging bigram.
In the remaining part of the paper we set $\lambda = 0.1$.
A lower lambda increases the number of additional phrase variations.
If the overhang to the left or right is more than two words we assume that the longer phrase is unique enough compared to the shorter phrase.

\paragraph{Redundant Shorter Candidates:}
Third, we discard shorter phrases that are already well represented by longer phrases.
Given a candidate phrase $p^i$, we determine its set $M^i$ of the shortest and distinct longer phrases among the candidates, i.e., any phrase $p^j \in M$ must not be a longer phrase of any other $p^l \in M$ and must be \emph{incompatible} with any other $p^l \in M$.
A phrase $p^j$ is \emph{incompatible} with $p^l$ if they share a common subsequence ($p^i$ in this case), but continue differently in either direction.
As an example, \textit{happy birthday} is incompatible with \textit{great birthday}, but not with \textit{birthday party}.
We remove the phrase $p_i$ if $s(p^i) - \sum_{j \in M^i}{s(p^j)} < s_k$.
For instance, we would discard \textit{day} if the candidates \textit{memorial day} and \textit{st patricks day} were already covering most occurrences of \textit{day}.

\vspace{10pt}

Most approaches try to find already good initial candidates so that they only need to rank these in the final step.
In contrast to this, our approach first collects candidates in a broader way, performs an initial ranking, and then reduces the set of keyphrases for the final ranking.
The advantage of this strategy is that there are much less restrictions regarding potential keyphrases.
The final list may contain phrases that start and/or end with stop words as well as complete sentences.
At the same time the second part of our pipeline ensures that we keep the number of redundant phrases low.

\begin{figure}
  \centering
  \includegraphics[width=\linewidth]{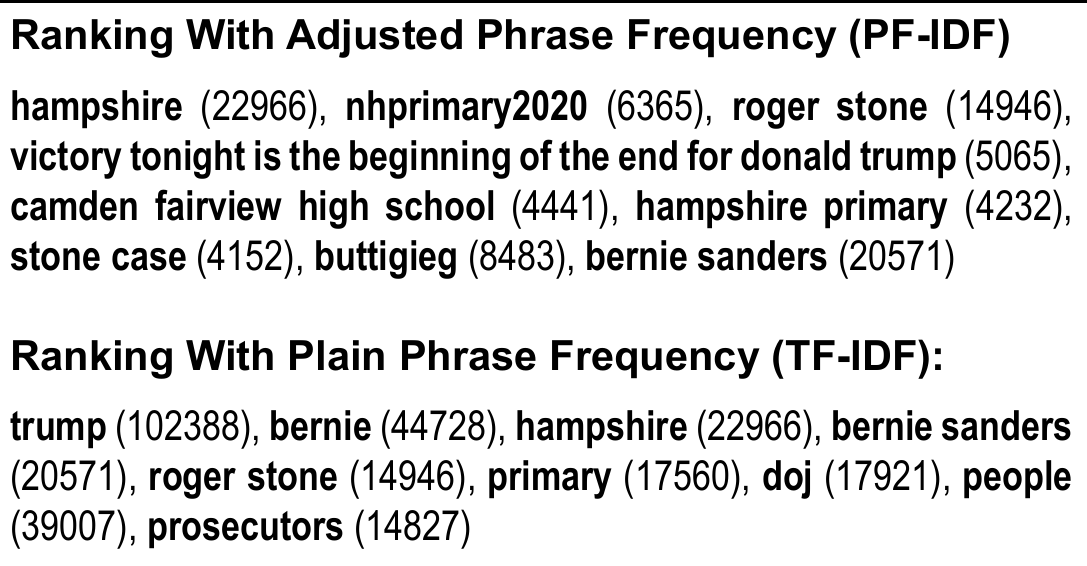}
  \caption{Top 10 keyphrases from 1m tweets published around Feb 12, 2020 (phrase frequency in brackets). The top depicts the final ranking using the adjusted phrase frequency (PF-IDF, $\mu \neq 1$), and the bottom using the plain phrase frequency (TF-IDF, $\mu = 1$)}
  \label{fig:example}
\end{figure}

Figure~\ref{fig:example} depicts an example of our approach applied to 1m tweets.
It shows that the top keyphrases contain both single terms and longer phrases, and that our sublinear scaling reduces the score of frequent terms that reveal little context.

\section{Evaluation}

\subsection{Performance Comparison}

\begin{table*}
\centering
\begin{tabular}{lrlrlrl}
\hline
                    & \multicolumn{2}{c}{\textbf{1k tweets (20k words)}} & \multicolumn{2}{c}{\textbf{100k tweets (2m words)}} & \multicolumn{2}{c}{\textbf{1m tweets (20m words)}} \\
                    & Time (s)                 & Speed-Up                & Time (s)                 & Speed-Up                 & Time (s)                 & Speed-Up                \\  \hline 
Baseline            & 0.896                    & 1                       & 49.16                    & 1                        & 642.28                    & 1                       \\
\textbf{Top k=100}  & 0.009                    & 100                      & 0.59                     & 83                       & 9.76                      & 66                      \\
\textbf{Top k=1000} & 0.110                    & 8                      & 1.33                     & 37                       & 15.79                     & 41                      \\ \hline
\end{tabular}
\caption{Performance of our candidate selection process compared to counting every \{1,...,m\}-gram (baseline).}
\label{tab:performance}
\end{table*}

We want to investigate the speed-up of our phrase candidate extraction pipeline (step 1 to 4) compared to the baseline, that is, extracting and calculating the PF-IDF score of every \{1,2,...,m\}-gram in the source.
To make the comparison fair, we disabled parallelization and used the same methods for both approaches to discard sub-phrases (step 3) and calculate the TF-IDF score (step 4), including the bigram-based index structure to quickly determine the document frequency in the reference collection.
We tested different configurations on a collection of lowercase tweets without punctuation.
We report the average duration of three runs.
The reference collection to determine the inverse document frequency comprises 2m tweets. Each tweet is made up of 20 words on average.
Table~\ref{tab:performance} shows that our selection process is between one and two orders of magnitude faster than counting every m-gram.
The run time of the second part of our pipeline is negligible compared to the selection process, it approximately needs between 200 and 400ms for the third case with 20m words.
We need to tokenize the input and convert it to a vector-based representation for both approaches.
This is comparable to the processing time of our $k = 100$ candidate selection process.
It should be noted that it typically takes even longer than our baseline to tag the same amount of data with a decent POS tagger.
For instance, the popular Stanford POS tagger would need approximately between half an hour and five hours to process 20m tokens, depending on the model~\cite{Horsmann2015}.

\subsection{Benchmarks}

\begin{table*}
\centering

\begingroup
\setlength{\tabcolsep}{4pt} 
\begin{tabular}{lrrrrrrrr}
\hline
	&	\multicolumn{2}{c}{\textbf{SemEval}} &	\multicolumn{2}{c}{\textbf{Krapivin}} &	\multicolumn{2}{c}{\textbf{Inspec}} &	\multicolumn{2}{c}{\textbf{NUS}}  \\
	& F\textsubscript{1}@5 & F\textsubscript{1}@10 & F\textsubscript{1}@5 & F\textsubscript{1}@10  & F\textsubscript{1}@5 & F\textsubscript{1}@10  & F\textsubscript{1}@5 & F\textsubscript{1}@10 \\ \hline
\textbf{ELSKE}					  & 22.0 & 22.5 & 22.6 & 19.5 & 20.7 & 23.6 & 26.1 & 20.6 \\ \hline
TF-IDF (POS)\textsuperscript{a)}  & 12.8 & 19.4 & 12.9 & 16.0 & 22.1 & 31.3 & 13.6 & 18.4 \\
TextRank\textsuperscript{a)}  	  & 17.6 & 18.7 & 18.9 & 16.2 & 22.3 & 28.1 & 19.5 & 19.6 \\
SingleRank\textsuperscript{a)}    & 13.5 & 17.6 & 18.9 & 16.2 & 21.4 & 30.6 & 14.0 & 17.3 \\
ExpandRank\textsuperscript{a)}    & 13.9 & 17.0 & 8.1  & 12.6 & 21.0 & 30.4 & 13.2 & 16.4 \\
TopicRank\textsuperscript{b)}     & 8.3  & 9.9  & 11.7 & 11.2 & n/a  & n/a  & 11.5 & 12.3 \\
YAKE\textsuperscript{c)}          & 15.1 & 21.2 & 21.5 & 19.6 & 20.4 & 22.3 & 15.9 & 19.6 \\ \hline
CopyRNN\textsuperscript{a)}  	  & 29.1 & 30.4 & 31.1 & 26.6 & 27.8 & 34.2 & 33.4 & 32.6 \\
CorrRNN\textsuperscript{b)}  	  & 32.0 & 32.0 & 31.8 & 27.8 & n/a  & n/a  & 35.8 & 33.0 \\ \hline

\end{tabular}
\caption{Benchmarks on present keyphrase prediction with reported values from a)~\cite{Meng2017} b)~\cite{Chen2020} and c)~\cite{martinc2020tntkid}. The last two RNN-based methods are supervised.}
\label{tab:benchmark}
\endgroup
\end{table*}

The goal of our method is to analyze collections of documents rather than single documents, but we still evaluated the performance on the SemEval~\cite{Kim2010},  Krapivin~\cite{Krapivin2008}, Inspec~\cite{Hulth2003}, and NUS~\cite{Nguyen2008} datasets to compare it with previous work.
We largely follow the procedure of Meng et al.~\cite{Meng2017}, which was also done in Chen et al.~\cite{Chen2020}.
We analyzed titles and abstracts, and measured the $F_1@k$ scores of present keyphrases of the gold standard.
We applied stemming when comparing the extracted keyphrases with the gold standard and for determining which keyphrases in the gold standard are \emph{present}, to make our results compatible with reported scores in related work.
We used the list of English stop words from the NLTK toolkit\footnote{https://gist.github.com/sebleier/554280}.
\textit{TF-IDF (POS)} describes the TF-IDF-based baseline method that uses POS-based rules for retrieving candidates~\cite{Hasan2010}.
The results in Table~\ref{tab:benchmark} show that while the \emph{supervised} recurrent neural network-based techniques take the lead, our approach is competitive among the unsupervised methods.

\section{Conclusion}

We presented a new technique for extracting keyphrases that exhibits several advantages which are particularly relevant if continuously incoming data has to be analyzed in a timely manner.
It can efficiently analyze large collections and imposes little restrictions on the length and structure of keyphrases, but it also performs reasonably well if targeted at individual documents.

\section*{Acknowledgments}

This research was supported by the German Science Foundation (DFG) as part of the project VAOST (project number 392087235) and as part of the Priority Program VA4VGI (SPP 1894).

\bibliographystyle{acl_natbib}
\bibliography{anthology,elske-paper}

\end{document}